\documentclass{appolb}
\usepackage{graphicx}
\usepackage[noadjust]{cite}
\usepackage{hyperref}
\usepackage{xcolor}
\usepackage{authblk}
\hypersetup{
    colorlinks,
    linkcolor={red!50!black},
    citecolor={blue!50!black},
    urlcolor={blue!80!black}
}

\begin{document}
\title{Four-lepton production from photon-induced reactions in pp collisions at the LHC%
}
\author[1,2]{M. Dyndal}
\author[3]{L. Schoeffel}
\affil[1]{AGH University of Science and Technology, al. Mickiewicza 30, 30-059 Krakow, Poland}
\affil[2]{Deutsches Elektronen-Synchrotron DESY, Notkestrasse 85, 22607 Hamburg, Germany}
\affil[3]{IRFU-SPP, CEA Saclay, 91191 Gif-sur-Yvette cedex, France}
\maketitle
\begin{abstract}
The cross sections for the reaction $\gamma\gamma\rightarrow 4\ell$ in proton--proton collisions are calculated at the LHC energies.
We show that the purely electroweak process $\gamma\gamma\rightarrow 4\ell$ can be studied at the LHC and can constitute a background to other processes with $4\ell$ or $2\ell$ final states.
\end{abstract}
  
\section{Introduction}
The study of photon--photon ($\gamma\gamma$) fusion processes in proton--proton ($pp$) collisions at the LHC provides a very important test of the electroweak theory~\cite{Chatrchyan:2011ci, Chatrchyan:2013foa, Aad:2015bwa}.
The $\gamma\gamma$ interactions leading to four leptons 
($\gamma\gamma \rightarrow 4\ell$, i.e. $\gamma\gamma \rightarrow 2e^+2e^-,$ $\gamma\gamma \rightarrow e^+e^-\mu^+\mu^-$ and $\gamma\gamma \rightarrow 2\mu^+2\mu^-$) can be another step towards these studies.

The baseline cross section, $\sigma_{\gamma\gamma \rightarrow 4\ell}$, has been  first estimated in the low-energy approximation~\cite{c1,c2,c3}. 
Since then, the theoretical computations of these processes have been extended to high energy domain, at center-of-mass energies of the $\gamma \gamma$ system around the TeV scale~\cite{c7,c8,c9,c10,Bredenstein:2004ef,c11, daSilva:2012pr}.
This becomes particularly interesting when these calculations are considered at the LHC for $pp$ collisions.

Two-photon production of four leptons can be formed via purely exclusive process: the leptons are produced  through the $\gamma\gamma$ interaction, where the photons are emitted coherently by each of the colliding protons.
In general, it is also possible that one or both protons dissociate into some hadronic state. 
Then, the reaction is not exclusive but inelastic (as protons dissociate),
while being induced by photons emitted from quarks in the proton.
This is why the label photon-induced production of four leptons is used in the following.

The main purpose of these studies, besides the cross section estimation, is to provide the relative contribution of two-photon production of four leptons compared to other processes at the LHC, e.g. to the $Z\rightarrow 4\ell$ production~\cite{Aad:2014wra,CMS:2012bw} or Higgs boson decaying to four leptons~\cite{Aad:2014eva,Chatrchyan:2012xdj}.
This type of photon-induced processes may appear as a background to the exclusive $\gamma\gamma\rightarrow W^+W^-$ production~\cite{Chatrchyan:2013foa}, which is also discussed.

\section{The $\gamma\gamma \rightarrow 4\ell$ cross section}
\label{sec2}

\begin{figure}[!htbp]
\centering
  \includegraphics[scale=0.42]{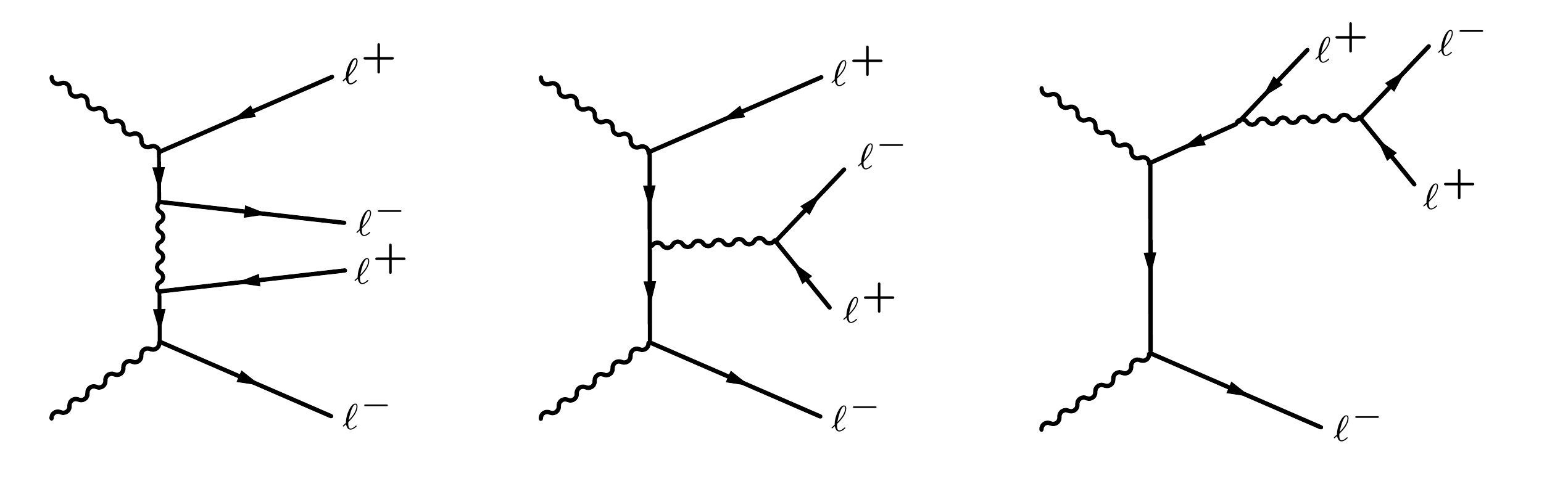}
  \put(-275,0){{(a)}}
  \put(-175,0){{(b)}}
  \put(-79,0){{(c)}}
   \caption[]{Examples of leading-order diagrams for the process $\gamma \gamma \rightarrow 4\ell$. All contributing diagrams form terms of $\mathcal{O}(\alpha_{em}^4)$, where $\alpha_{em}$ is the electromagnetic coupling constant.}
  \label{FIGdiag1}
\end{figure}

The elementary process cross section calculation is based 
essentially on summation of the leading-order amplitudes, represented by their 
Feynman diagrams in Figure~\ref{FIGdiag1}.
From these diagrams, it can be shown that  
the bulk of the contribution comes from the production (a):
the two photons radiate a pair of leptons and two of these leptons undergo scattering (the $t$-channel photon exchange) with its characteristic
singularity in the forward direction (when $t$ tends towards zero). 
In general, this diagram is divergent in the limit of massless leptons, which means the leptons are emitted almost collinear to the beams.

It is also worth noticing that these calculations do not include $\gamma\gamma\rightarrow ZZ \rightarrow 4\ell$ process~\cite{Jikia}, which is forbidden at the tree-level, hence this contribution is highly suppressed and can be neglected in these studies.

The calculations of the elementary cross section
$\sigma_{\gamma \gamma \rightarrow 4\ell}$ are performed
using full tree-level Monte Carlo (MC) program {\sc Madgraph5$\_$aMC$@$NLO}~\cite{Alwall:2014hca}, referred to hereafter as {\sc Madgraph}.
Indeed, it is important to recall that the cross sections for the two-photon production of four leptons using {\sc Madgraph} have already been shown to be in a good agreement with independent calculations~\cite{Bredenstein:2004ef}.

\begin{figure}[!htbp]
\centering
  \includegraphics[scale=0.34]{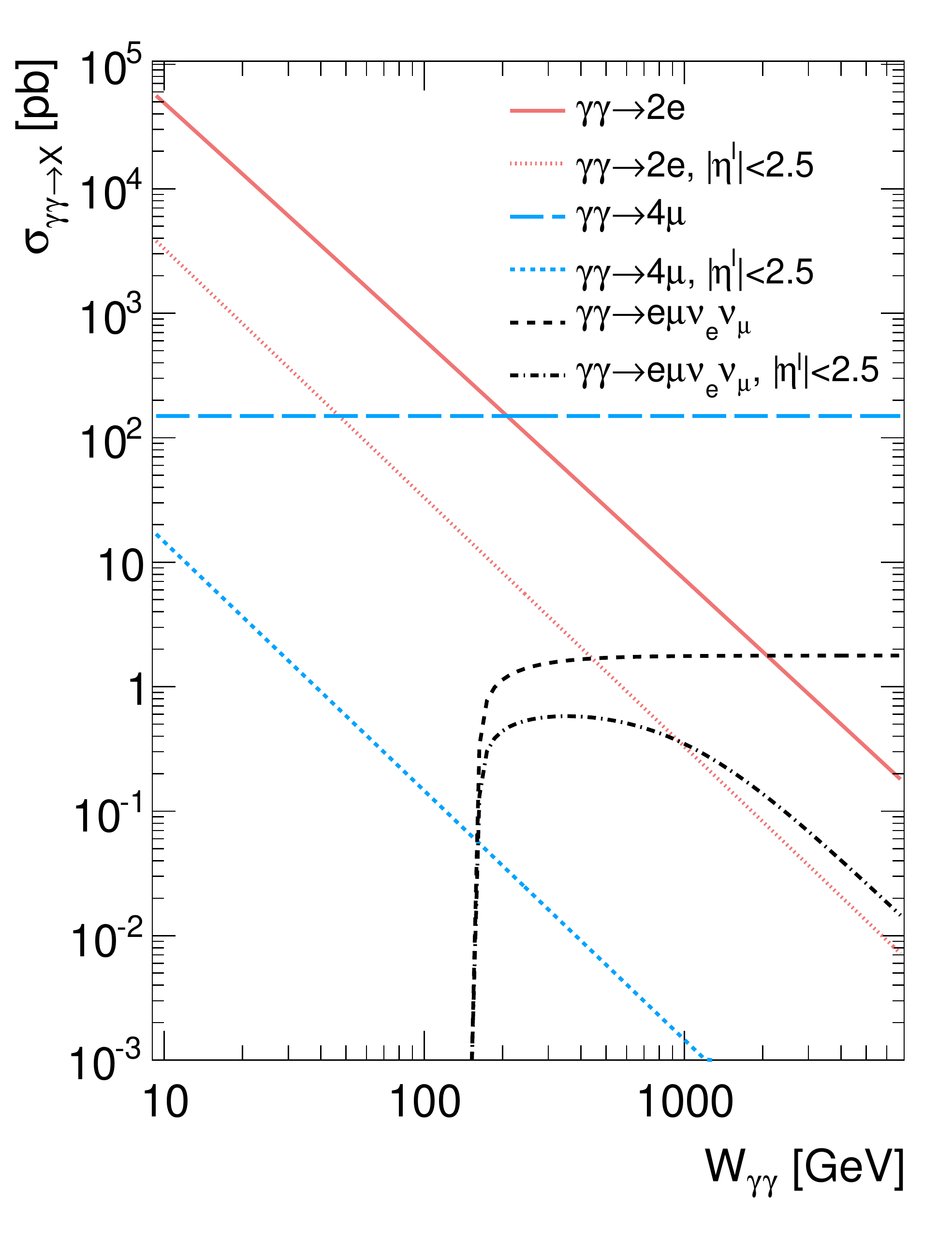}
   \caption[]{Leading-order cross sections for the elementary processes $\gamma \gamma \rightarrow X$ as a function of the 
energy of the two-photon system, $W_{\gamma \gamma}$.
Different electroweak final states are pictured with and without requirements on the pseudorapidities of leptons ($|\eta^{\ell}|<2.5$).
The cross section for lepton-neutrino pair production involves decays of $W$ bosons. }
  \label{FIGelem_xs_comp1}
\end{figure}

It is interesting to compare the typical values of electroweak cross sections that occur in $\gamma\gamma$ collisions up to energies accessible at the LHC.
This is illustrated in Figure~\ref{FIGelem_xs_comp1}.
The cross sections for $\gamma \gamma \rightarrow X$ are presented as a function of the energy of the two-photon system ($W_{\gamma \gamma}$) for different final states:
four leptons, two leptons and two $W$ bosons of opposite charges decaying to leptons and neutrinos.
Prior to any angular cuts, Figure~\ref{FIGelem_xs_comp1} shows that the cross section for the process $\gamma \gamma \rightarrow 2 \ell$
is behaving as $1/W_{\gamma \gamma}^2$, while the cross section for $\gamma \gamma \rightarrow 4 \ell$
is constant with $W_{\gamma \gamma}$.
The same constant behavior is obtained for the production of two $W$ bosons once the production threshold is reached.
In each case this constant behavior as a function of $W_{\gamma \gamma}$
is due to the spin-1 $t$-channel particle exchange (photon or $W$ boson respectively), which implies that the cross section  is completely dominated by configurations where the leptons are emitted in very forward directions, almost collinear to the beams.
Then, no dependence on the energy of the two-photon system is expected.

However, we are interested in the possibility of measuring such processes in collider experiments. 
Therefore, a more realistic view is obtained for the typical values of cross sections for $\gamma \gamma \rightarrow X$,
once some angular cuts are applied on the leptons, in order to mimic their acceptances inside the trackers of e.g. ATLAS or CMS experiments at the LHC~\cite{Aad:2008zzm,Chatrchyan:2008aa} (see Figure~\ref{FIGelem_xs_comp1}).
In this case the cross sections for $\gamma \gamma \rightarrow 4\ell$ and $\gamma \gamma \rightarrow W^+W^-$ reactions behave as $1/W_{\gamma \gamma}^2$.
Again, this is due to the $t$($u$)-channel poles in the
forward (backward) directions: with increasing energy, these cross sections increase in the very forward and backward direction being proportional to $W_{\gamma \gamma}^2$
but they decrease in the central region, being proportional to $1/W_{\gamma \gamma}^2$.

\section{Photon-induced production of four leptons at the LHC}
\label{sec3}

In order to compute the  cross section for the photon-induced production of four leptons in $pp$ collisions at the LHC,
at $pp$ center-of-mass energy $\sqrt{s}$, we follow
a similar methodology as presented in Refs.~\cite{Ball:2013hta, Bierweiler:2012kw, Luszczak:2014mta, daSilveira:2015hha} for the photon-induced $W^+W^-$ production. 
At first what is needed is the calculation of 
the elementary cross section of the $\gamma\gamma$ interactions leading to four leptons, $\sigma_{\gamma \gamma \rightarrow 4\ell}$, performed in Section~\ref{sec2}.

For the exclusive reaction, each of the two incoming protons emits a quasi-real photon which then fuse to give a centrally produced four-lepton final state ($\gamma\gamma \rightarrow 4\ell$). The calculations rely on the so-called  equivalent photon approximation (EPA) \cite{fermi,ww,t1,t2,Budnev:1973tz}.
For the modelling of the exclusive reaction in $pp$ collisions, the EPA is used with the standard dipole parameterization of proton electromagnetic form-factors \cite{Ernst:1960zza}.
No specific corrections are applied to take into account proton absorptive effects~\cite{Dyndal:2014yea}.
 
When one or both protons dissociate into a hadronic state, dedicated parton distribution functions (PDFs) for photons are used~\cite{Ball:2013hta, Martin:2004dh}. 
In view of making realistic estimates for what can be observed at the LHC,
another interest of using {\sc Madgraph} is that it allows to take into account QCD effects, such as the underlying event and parton shower in single and double-dissociative events. For this purpose, {\sc Madgraph} is 
interfaced with {\sc Pythia~8}~\cite{Sjostrand:2007gs}.

\begin{figure}[!t]
\centering
  \includegraphics[scale=0.5]{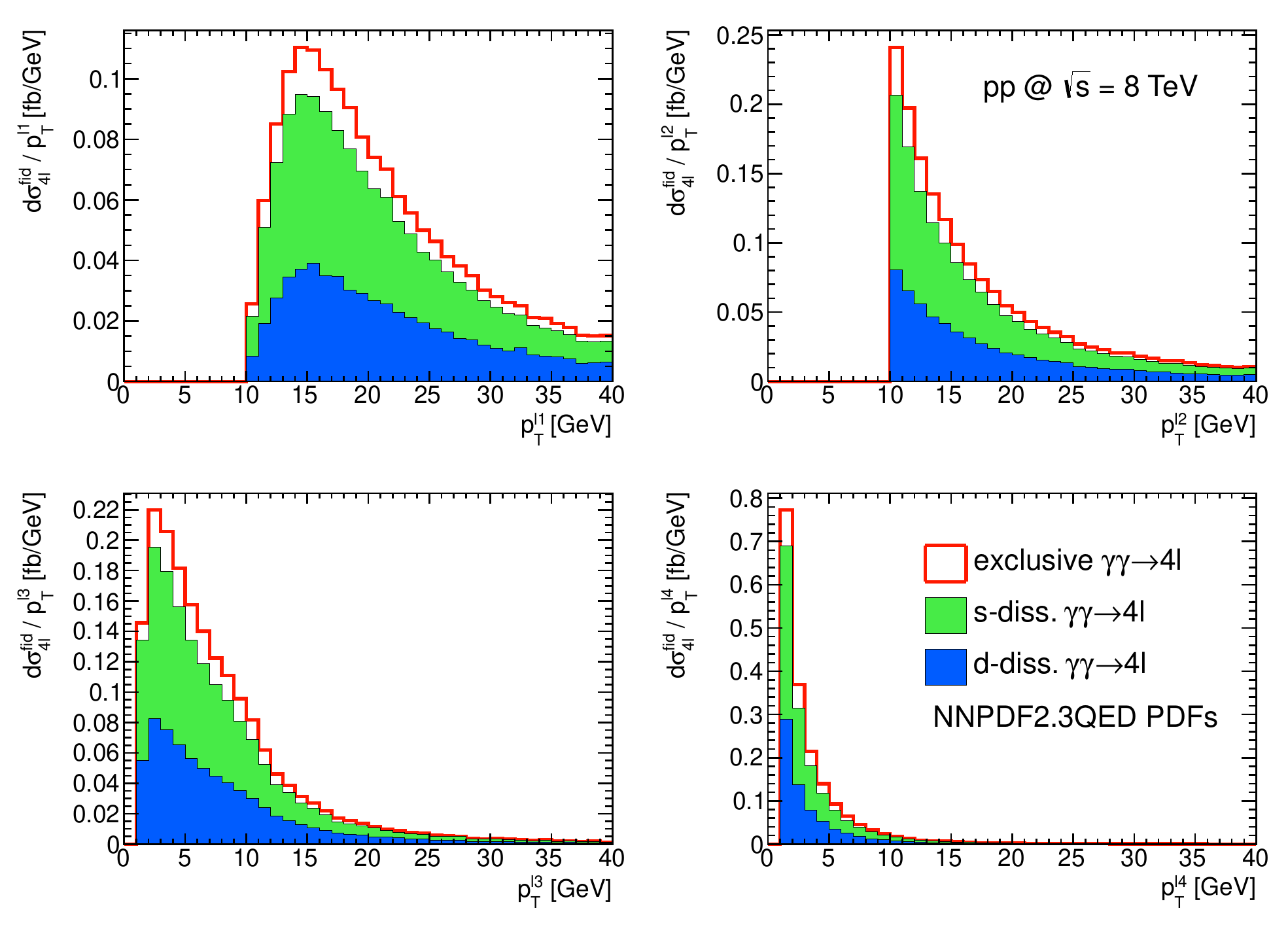}
\put(-215,102){{(a)}}
\put(-70,102){{(b)}}
\put(-215,0){{(c)}}
\put(-70,0){{(d)}}
   \caption[]{Differential cross sections for $\gamma\gamma \rightarrow 4\ell ~(\ell=e,~\mu)$ production in $pp$ collisions at $\sqrt{s} = 8$ TeV as a function of (a) leading lepton $p_\mathrm{T}$, (b) sub-leading lepton $p_\mathrm{T}$, (c) third lepton $p_\mathrm{T}$ and (d) fourth lepton $p_\mathrm{T}$. The stacked histograms, in top-to-bottom order, represent the exclusive, single-dissociative and double-dissociative contributions.
Some fiducial cuts are applied (see text).}
  \label{FIGpt}
\end{figure}
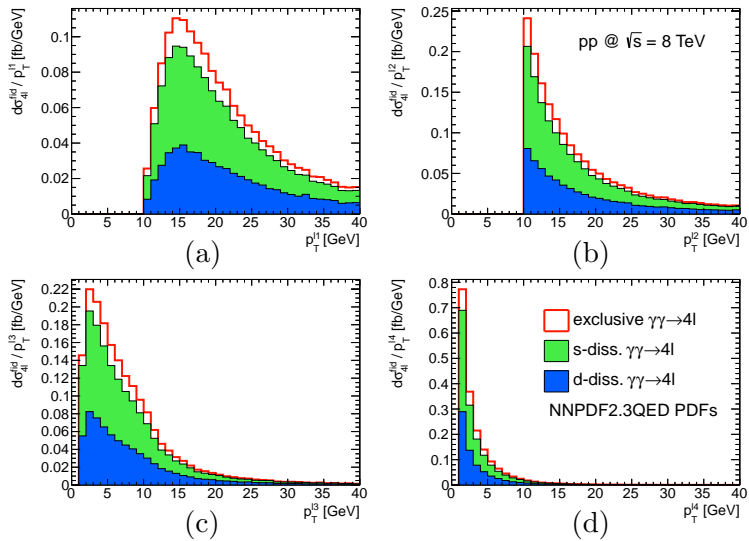

Differential cross sections for $\gamma\gamma \rightarrow 4\ell$ production in $pp$ collisions at $\sqrt{s} = 8$ TeV as a function
of the transverse momenta for all outgoing leptons are presented in Figure~\ref{FIGpt}, where the NNPDF2.3QED PDFs are used for the proton-dissociative case.
In this figure, some kinematic requirements are imposed to mimic realistic measurements in the ATLAS and CMS experiments.
As in Figure~\ref{FIGelem_xs_comp1},
the absolute values of the pseudorapidities for all leptons are required to be $|\eta^{\ell}|<2.5$.
In addition, the transverse momenta for the leading and sub-leading leptons are taken to be above $10$ GeV.
This requirement is imposed to ensure high triggering efficiency in the ATLAS and CMS experiments during nominal LHC runs~\cite{Aad:2014rra, Khachatryan:2015hwa}.
The two other leptons are requested to have transverse momenta above $1$ GeV. This threshold should be taken as low as possible, in the limit where it is still possible to use trackers in order to discriminate between e.g. electrons and pions~\cite{ATLAS-CONF-2011-128}.
Finally, the angular separation between each pair of leptons, $\Delta R_{\ell\ell} = \sqrt{\Delta \phi_{\ell\ell}^2 + \Delta \eta_{\ell\ell}^2}$, is taken to be $\Delta R_{\ell\ell} > 0.1$ to allow the experimental separation of these objects.

There are two important points to observe in Figure~\ref{FIGpt}: (1)  the single and double-dissociative reactions have similar cross sections in the kinematic phase space defined above and (2) this cross section (summed) is about ten times larger than for the exclusive case.
The exclusive and proton-dissociative contributions can be easily separated using lepton kinematic distributions, like total $p_\mathrm{T}$ of the $4\ell$ system~\cite{Chatrchyan:2011ci, Aad:2015bwa}, since the lepton directions and momenta can be precisely measured.

Comparison of total cross sections for $\gamma\gamma \rightarrow 4\ell$ production in $pp$ collisions at $\sqrt{s} = 8$ TeV and 13 TeV with typical lepton kinematic cuts is shown in Table~\ref{table1}.
Here also the two different photon PDFs are compared for modelling proton-dissociative contributions.
The double-dissociative part is about 3 times larger when using MRST2004QED PDF set. This is due to the growth of the MRST2004QED photon PDFs over the NNPDF2.3QED set at smaller $x$ values of photons ($x< 0.005$)~\cite{Ball:2013hta}.

The cross sections compared in Table~\ref{table1} are of the order of $\sim$fb, therefore possible to be measured at the LHC under nominal running conditions. 
The identification of $\gamma\gamma\rightarrow 4\ell$ events of interest can be performed using veto on additional charged-particle tracks, successfully used by the ATLAS and CMS experiments to significantly reduce the inclusive background~\cite{Chatrchyan:2011ci, Chatrchyan:2013foa, Aad:2015bwa}.
Moreover, all quoted cross sections can be increased by a factor of 2.3 when the requirement on leading and subleading lepton $p_\mathrm{T}$ is lowered from 10 GeV to 5 GeV.

\begin{table}
\centering
\caption[]{
Total cross sections for $\gamma\gamma\rightarrow 4\ell$~($\ell=e,~\mu$) production in $pp$ collisions for $\sqrt{s}=8$ TeV and 13 TeV. Leptons with $|\eta^{\ell}|<2.5$ and $p_\mathrm{T}^{\ell1}>10$ GeV, $p_\mathrm{T}^{\ell2}>10$ GeV, $p_\mathrm{T}^{\ell3}>1$ GeV, $p_\mathrm{T}^{\ell4}>1$ GeV are considered. Different contributions are presented: exclusive, single-dissociative and double-dissociative. For modelling of single and double-dissociative contributions, two different photon PDFs are compared.
}
\begin{tabular}{lccc}
  \hline \hline  
  ${\sqrt s}$ & $\sigma^\mathrm{excl}_{\gamma \gamma \rightarrow 4 \ell}$ & 
$\sigma^\mathrm{sdiss}_{\gamma \gamma \rightarrow 4 \ell}$  {\footnotesize with} &
$\sigma^\mathrm{ddiss}_{\gamma \gamma \rightarrow 4 \ell}$  {\footnotesize with}\\
 &   &  {\scriptsize NNPDF2.3 (MRST2004) QED} & {\scriptsize NNPDF2.3 (MRST2004) QED} \\
  \hline
  $8$  TeV & $0.22$ fb & $0.91$ ($1.4$) fb & $0.71$ ($2.0$) fb \\
  $13$ TeV & $0.29$ fb & $1.2$ ($1.8$) fb  & $0.86$ ($2.7$) fb \\
  \hline \hline
\end{tabular}
\label{table1}
\end{table}

\section{Background to inclusive production of four leptons at the LHC}
\label{sec4}

The inclusive production of four leptons has been recently measured by the ATLAS experiment~\cite{Aad:2015rka}.
Let us recall that in the Standard Model, the $4\ell$ production in the invariant mass range domain around the $Z$  boson resonance
occurs dominantly via the $s$-channel process, where the $Z$ boson decaying to  leptons includes the production of an additional
lepton pair from the internal conversion of a virtual $Z$ boson or a virtual photon. 
Also, the cross section is significantly increased when both $Z$ bosons are produced on-shell, resulting in the rise in the $m_{4\ell}$ spectrum observed around $180$ GeV.
In addition to the $q{\bar q}$-initiated process of $4\ell$ production, there are some  gluon--gluon initiated processes, 
$gg \rightarrow 4\ell$, needed to be considered~\cite{Aad:2014wra,CMS:2012bw}.
The gluon-induced processes significantly contribute to 
the on-shell Higgs boson production that obviously gives a sharp resonance around $125$ GeV~\cite{Aad:2014eva,Chatrchyan:2012xdj}.

\begin{figure}[!htbp]
\centering
  \includegraphics[scale=0.4]{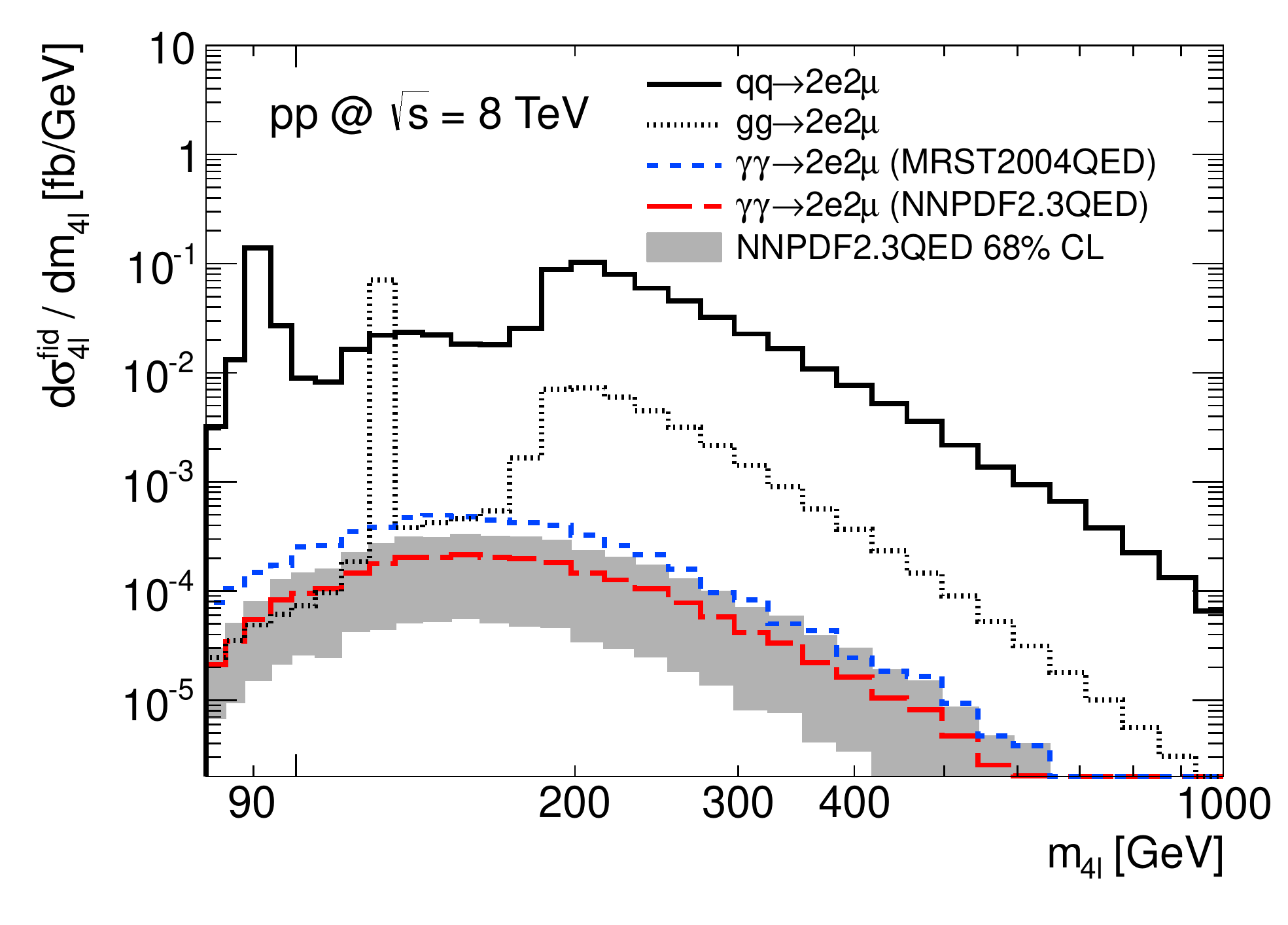}
   \caption[]{Differential cross sections (d$\sigma$/d$m_{4\ell}$) from the $\gamma\gamma$, $q\bar{q}$ and $gg$ initial states at $\sqrt{s} = 8$ TeV for the $e^+e^-\mu^+\mu^-$ final state in the fiducial phase space from Ref.~\cite{Aad:2015rka}. 
For $\gamma\gamma$ production, two different photon PDFs are compared, where the shaded region indicates 68\% confidence level for the NNPDF2.3QED PDFs. }
  \label{FIGinvmassglobal}
\end{figure}

A comparison of photon-induced $4\ell$ contribution to the abovementioned processes is presented in Figure~\ref{FIGinvmassglobal}, alternatively for MRST2004QED and NNPDF2.3QED PDFs, where the latter includes additional PDFs uncertainty estimation.
For this figure, the same fiducial cuts as in~\cite{Aad:2015rka} are applied.
Similarly as in Ref.~\cite{Aad:2015rka}, the $q\bar{q}\rightarrow 4\ell$ and on-shell Higgs-boson production contributions are modeled at next-to-leading-order (in perturbative QCD) using {\sc Powheg-Box}~\cite{Melia:2011tj} MC generator. 
The non-resonant $gg\rightarrow 4\ell$ process is simulated at leading-order with {\sc MCFM}~\cite{Campbell:2013una} program. Both generators are interfaced with CT10~\cite{Lai:2010vv} PDF set.

Figure~\ref{FIGinvmassglobal} shows that, depending on the photon PDFs used, the photon-induced part can reach up to 5\% of the standard $q\bar{q}$ contribution 
in the non-resonant mass range of the $Z$ boson (i.e. $70$ GeV $< m_{4\ell} < 80$ GeV and $100$ GeV $< m_{4\ell} < 110$ GeV) 
and up to 3\% in the mass range of the Higgs boson.

\section{Background to exclusive $\gamma\gamma\rightarrow W^+W^-$ production at the LHC}
\label{sec5}
The enhancement of the elementary $\gamma\gamma\rightarrow 4\ell$ cross section at large lepton pseudorapidities gives the possibility that some of the four leptons will escape detection, whereas other could mimic the exclusive 2$\ell$ final state.
In particular, this is interesting when comparing to processes with relatively small cross sections at the LHC, like the exclusive $\gamma\gamma\rightarrow W^+W^-$ production, used to test the anomalous $\gamma\gamma WW$ couplings~\cite{Chatrchyan:2013foa,Belanger:1992qi}.
Here the $e^{\pm}\mu^{\mp}\nu_e\nu_{\mu}$ final state is examined, which  forms effectively an experimental state with two charged-leptons.

Using the same fiducial region definition as in Ref.~\cite{Chatrchyan:2013foa} ($p_\mathrm{T}^{\ell}>20$ GeV and $|\eta^{\ell}|<2.4$) applied to one of the $e^{\pm}\mu^{\mp}$ pair in the  $e^+e^-\mu^+\mu^-$ final state and imposing additional pseudorapidity veto ($|\eta^{\ell}|>2.4$) on the remaining $e^{\mp}\mu^{\pm}$ pair, we obtain:
\begin{equation}
\sigma_{pp(\gamma\gamma)\rightarrow pp e^+e^-\mu^+\mu^-}^\mathrm{fid} = 0.4~\mathrm{fb}~, \nonumber
\end{equation}
to be compared with~\cite{Chatrchyan:2013foa}:
\begin{equation}
\sigma_{pp(\gamma\gamma)\rightarrow pp W^+W^-\rightarrow pp e^{\pm}\mu^{\mp}\nu_e\nu_{\mu}}^\mathrm{fid} = 0.66~\mathrm{fb}~. \nonumber
\end{equation}

However, imposing final requirement on the $e^{\pm}\mu^{\mp}$ pair transverse momentum, $p_\mathrm{T}^{e^{\pm}\mu^{\mp}} > 30$ GeV~\cite{Chatrchyan:2013foa}, leads to:
\begin{equation}
\sigma_{pp(\gamma\gamma)\rightarrow e^+e^-\mu^+\mu^-X}^\mathrm{fid, cut} = 0.03~\mathrm{fb}~. \nonumber
\end{equation}

It is worth mentioning that the contribution of the process  $\gamma\gamma\rightarrow e^+e^-\mu^+\mu^-$
can be estimated using same-sign lepton pairs, i.e. $e^{\pm}\mu^{\pm}$ whereas the remaining $e^{\mp}\mu^{\mp}$ pair is emitted at large pseudorapidities. 
This would allow to reduce the contamination from other (inclusive and exclusive) processes, which mostly produce opposite-sign lepton pair final states.

\section{Conclusion and outlook}

The production of $\gamma\gamma\rightarrow 4\ell$ events at the LHC provides an ideal testing ground for the study of QED at $\mathcal{O}(\alpha_{em}^4)$ elementary process level.
The obtained values of the cross sections show that it can be possible to measure  the $\gamma\gamma\rightarrow 4\ell$ reactions in $pp$ collisions at the LHC.
Indeed, it has been already demonstrated that the $\gamma \gamma$ interactions can be studied at the LHC, under normal running conditions, i.e. in the presence of additional pile-up events that spoil the identification of the $\gamma \gamma$ interactions of interest.

For $4\ell$ production in $pp$ collisions, the mechanism of two-photon fusion is compared with that of quark--antiquark and gluon--gluon. 
The predictions show that the total $\gamma\gamma$ cross section should contribute up to 5\% in some kinematical regions accessible by the LHC experiments.
Moreover, the $e^+e^-\mu^+\mu^-$ final state with one of the $e^{\pm}\mu^{\mp}$ pair emitted at large pseudorapidities can be a background to the $\gamma\gamma\rightarrow W^+W^-\rightarrow e^{\pm}\mu^{\mp}\nu_e\nu_{\mu}$ reaction, reducing the sensitivity for anomalous $\gamma\gamma WW$ coupling searches.
Finally, the photon-induced production of $4\ell$ system can also be a non-reducible background in searches for anomalous $\gamma \gamma ZZ$ couplings~\cite{Belanger:1992qi} at the LHC, especially with dedicated proton tagging~\cite{Chapon:2009hh, Gupta:2011be}.

\section*{Acknowledgements}
We thank David d'Enterria for useful discussions. This work was partly supported by the Polish National Science Center under contract No. UMO-2012/05/B/ST2/02480.



\begin{thebibliography}{unsrt}


\bibitem{Chatrchyan:2011ci}
  S.~Chatrchyan {\it et al.}  [CMS Collaboration],
  JHEP {\bf 1201} (2012) 052.

\bibitem{Chatrchyan:2013foa}
  S.~Chatrchyan {\it et al.}  [CMS Collaboration],
  JHEP {\bf 1307} (2013) 116.


\bibitem{Aad:2015bwa}
  G.~Aad {\it et al.} [ATLAS Collaboration],
  Phys.\ Lett.\ B {\bf 749} (2015) 242.

\bibitem{c1}
  L.N. Lipatov and G.V. Frolov, 
  Pisma Zh.\ Eksp.\ Teor.\ Fiz.\  {\bf 10}, 399 (1969); 
  JETP Lett.  {\bf 10} 254 (1969).

\bibitem{c2}
 V.G. Serbo, 
 Pisma Zh.\ Eksp.\ Teor.\ Fiz.\  {\bf 12}, 33 (1970).

\bibitem{c3}
 H. Cheng and T.T.Wu, Phys. Rev.  D {\bf 1} (1970) 3414.

\bibitem{c7}
P. De Causmaecker, R. Gastmans, W. Troost, Tai Tsun Wu, Phys.
Lett.  {\bf B105} (1981) 215.

\bibitem{c8}
P. De Causmaecker, R. Gastmans, W. Troost, Tai Tsun Wu, Nucl.
Phys.  B {\bf 206} (1982) 53.

\bibitem{c9}
F.A. Berends {\it et al.}, Nucl. Phys. B {\bf 206} (1982) 61.

\bibitem{c10}
S. Dittmaier,  {\it Weyl-van-der-Waerden formalism for
helicity amplitudes of massive particles}, hep-ph/9805445 (1998).

\bibitem{Bredenstein:2004ef}
  A.~Bredenstein, S.~Dittmaier and M.~Roth,
  Eur.\ Phys.\ J.\ C {\bf 36} (2004) 341. See Tables 2 and 3 for a comparison of independent calculations with {\sc Madgraph}.

\bibitem{c11}
T.Shishkina and I.Sotsky, 
{\it Production of two electron-positron couples in electroweak gamma-gamma interaction}, hep-ph/0410010 (2004).

\bibitem{daSilva:2012pr} 
  W.~da Silva and F.~Kapusta,
  Phys.\ Lett.\ B {\bf 718}, 577 (2013).



\bibitem{Aad:2014wra}
  G.~Aad {\it et al.} [ATLAS Collaboration],
  Phys.\ Rev.\ Lett.\  {\bf 112}, no. 23, 231806 (2014).

\bibitem{CMS:2012bw}
  S.~Chatrchyan {\it et al.} [CMS Collaboration],
  JHEP {\bf 1212} (2012) 034.

\bibitem{Aad:2014eva}
  G.~Aad {\it et al.} [ATLAS Collaboration],
  Phys.\ Rev.\ D {\bf 91}, no. 1, 012006 (2015);
  Phys.\ Lett.\ B {\bf 716} (2013) 1.

\bibitem{Chatrchyan:2012xdj}
  S.~Chatrchyan {\it et al.} [CMS Collaboration],
  Phys.\ Lett.\ B {\bf 716} (2012) 30.

\bibitem{Jikia}
  G.V. Jikia, 
  Phys. Lett. {\bf B298} (1993) 224; Nucl. Phys. {\bf B405} (1993) 24.

\bibitem{Alwall:2014hca}
  J.~Alwall {\it et al.},
  JHEP {\bf 1407} (2014) 079.

\bibitem{Aad:2008zzm} 
  G.~Aad {\it et al.} [ATLAS Collaboration],
  JINST {\bf 3}, S08003 (2008).
  
  \bibitem{Chatrchyan:2008aa} 
  S.~Chatrchyan {\it et al.} [CMS Collaboration],
  JINST {\bf 3}, S08004 (2008).





\bibitem{Ball:2013hta}
  R.~D.~Ball {\it et al.} [NNPDF Collaboration],
  Nucl.\ Phys.\ B {\bf 877} (2013) 290.
  
  

\bibitem{Bierweiler:2012kw} 
  A.~Bierweiler, T.~Kasprzik, J.~H.~Kühn and S.~Uccirati,
  JHEP {\bf 1211}, 093 (2012).

\bibitem{Luszczak:2014mta} 
  M.~Luszczak, A.~Szczurek and C.~Royon,
  JHEP {\bf 1502}, 098 (2015).
  

\bibitem{daSilveira:2015hha} 
  G.~G.~da Silveira and V.~P.~Goncalves,
  Phys.\ Rev.\ D {\bf 92}, no. 1, 014013 (2015).
  

\bibitem{fermi}
  C.~F.~von Weizsacker,
  Z.\ Phys.\  {\bf 88}, 612 (1934).


\bibitem{ww}
E.~Williams, {Phys. Rev. {\bf 45} (1934)
  729}.

\bibitem{t1}
  M.~S.~Chen, I.~J.~Muzinich, H.~Terazawa and T.~P.~Cheng,
  Phys.\ Rev.\ D {\bf 7} (1973) 3485.

\bibitem{t2}
  H.~Terazawa,
  Rev.\ Mod.\ Phys.\  {\bf 45} (1973) 615.

\bibitem{Budnev:1973tz}
V.~Budnev, I.~Ginzburg, G.~Meledin, and V.~Serbo,  {Nucl. Phys. B {\bf 63}
  (1973)  519}.


\bibitem{Ernst:1960zza}
  F.~J.~Ernst, R.~G.~Sachs and K.~C.~Wali,
  Phys.\ Rev.\  {\bf 119} (1960) 1105.
  

\bibitem{Dyndal:2014yea}
  M.~Dyndal and L.~Schoeffel,
  Phys.\ Lett.\ B {\bf 741} (2015) 66.




\bibitem{Martin:2004dh}
  A.~D.~Martin, R.~G.~Roberts, W.~J.~Stirling and R.~S.~Thorne,
  Eur.\ Phys.\ J.\ C {\bf 39} (2005) 155.






\bibitem{Sjostrand:2007gs}
  T.~Sjostrand, S.~Mrenna and P.~Z.~Skands,
  Comput.\ Phys.\ Commun.\  {\bf 178} (2008) 852.


\bibitem{Aad:2014rra} 
  G.~Aad {\it et al.} [ATLAS Collaboration],
  Eur.\ Phys.\ J.\ C {\bf 74}, no. 7, 2941 (2014); Eur.\ Phys.\ J.\ C {\bf 74}, no. 11, 3130 (2014).


\bibitem{Khachatryan:2015hwa} 
  V.~Khachatryan {\it et al.} [CMS Collaboration],
  JINST {\bf 8}, P11002 (2013); JINST {\bf 10}, no. 06, P06005 (2015). 


\bibitem{ATLAS-CONF-2011-128}
     G.~Aad {\it et al.} [ATLAS Collaboration],
     {\it Particle Identification Performance of the ATLAS
                       Transition Radiation Tracker}, 
     ATLAS-CONF-2011-128, cds.cern.ch/record/1383793.
     




\bibitem{Aad:2015rka} 
  G.~Aad {\it et al.} [ATLAS Collaboration],
  Phys.\ Lett.\ B {\bf 753}, 552 (2016).




\bibitem{Melia:2011tj} 
  T.~Melia, P.~Nason, R.~Rontsch and G.~Zanderighi,
  JHEP {\bf 1111}, 078 (2011).

\bibitem{Campbell:2013una} 
  J.~M.~Campbell, R.~K.~Ellis and C.~Williams,
  JHEP {\bf 1404}, 060 (2014).
  
\bibitem{Lai:2010vv} 
  H.~L.~Lai, M.~Guzzi, J.~Huston, Z.~Li, P.~M.~Nadolsky, J.~Pumplin and C.-P.~Yuan,
  Phys.\ Rev.\ D {\bf 82}, 074024 (2010).




\bibitem{Belanger:1992qi} 
  G.~Belanger and F.~Boudjema,
  Phys.\ Lett.\ B {\bf 288}, 210 (1992).

\bibitem{Chapon:2009hh} 
  E.~Chapon, C.~Royon and O.~Kepka,
  Phys.\ Rev.\ D {\bf 81}, 074003 (2010).

\bibitem{Gupta:2011be} 
  R.~S.~Gupta,
  Phys.\ Rev.\ D {\bf 85}, 014006 (2012).

  

\end{thebibliography}
\end{document}